\documentclass[aps,prb,twocolumn,superscriptaddress,showpacs,amsmath,amssymb,longbibliography]{revtex4-2}
\usepackage[english]{babel}
\usepackage{amsmath}
\usepackage{bm}
\usepackage{graphicx,bbm} 
\usepackage{times}
\usepackage{epsfig} 
\usepackage{soul}
\usepackage[utf8]{inputenc}
\usepackage[colorlinks,linkcolor=blue,citecolor=blue,urlcolor=blue]{hyperref}
\usepackage{color}
\usepackage{latexsym}
\usepackage{ulem}
\definecolor{nred} {RGB}{224,0,0}
\definecolor{nblue} {RGB}{28,130,185}
\definecolor{dgreen} {RGB}{78,138,21}
\definecolor{norange}{RGB}{230,120,20}

\begin{document} 
\title{Ballistic transport in integrable quantum lattice models with degenerate spectra}
\author{M. Mierzejewski}
\affiliation{Department of Theoretical Physics, Faculty of Fundamental Problems of Technology, Wroc\l aw University of Science and Technology, 50-370 Wroc\l aw, Poland}
\author{J. Herbrych}
\affiliation{Department of Theoretical Physics, Faculty of Fundamental Problems of Technology, Wroc\l aw University of Science and Technology, 50-370 Wroc\l aw, Poland}
\author{P. Prelov\v{s}ek}
\affiliation{Jo\v zef Stefan Institute, SI-1000 Ljubljana, Slovenia}
\affiliation{Faculty of Mathematics and Physics, University of Ljubljana, SI-1000 Ljubljana, Slovenia}

\date{\today}
\begin{abstract}
We study the ballistic transport in integrable quantum lattice models, i.e., in the spin XXZ and Hubbard chains, close to the noninteracting limit. t is by now well established that the stiffnesses of spin and charge currents reveal, at high temperatures, a discontinuous reduction (jump) when the interaction is introduced. We show that the jumps are related to the large degeneracy of the parent noninteracting models and, more generally, can appear in other integrable models with  macroscopic degeneracies. These degeneracies are properly captured by the degenerate perturbation calculations which may be performed for large systems. We find that the discontinuities and the quasilocality of the conserved current in this limit can be traced back to  the nonlocal character of an effective interaction. From the latter observation we identify a class of observables which show discontinuities.
\end{abstract}

\maketitle

\section {Introduction} 
It is by now well established that the transport in integrable quantum lattice models is anomalous (for a recent review see \cite{bertini20}) and can remain dissipationless/ballistic even at infinite temperature $T$. The main interest focused on the conductivities and related ballistic components - the stiffnesses \cite{castella95,zotos96,zotos97,zotos99,hmeisner03,benz2005,hmeisner07,shastry2008,Sirker2009,herbrych11,
znidaric11,prosen11,prosen13,prosen14,pereira14,steinigeweg14,
ilievski17,ilievski171,karrasch17,bulchandani18,urichuk19} in the one-dimensional (1D) integrable models, mainly the spin XXZ and the Hubbard chains. Non-vanishing stiffnesses have been demonstrated via the sensitivity to the flux \cite{castella95} and via the Mazur inequality \cite{mazur,zotos97} that directly accounts for the conservation laws specific for integrable systems, i.e., the local conserved charges
 \cite{grabowski95}. The latter approach demonstrates that the ballistic transport can be recognized as a manifestation of non-thermal steady states \cite{rigol2009,Cassidy2011,vidmar2013,caux13,wouters14,pereira14,mestyan15,alba16,rigol16,vidmar16}. While the numerical evidence within the XXZ model at half-filling clearly reveals finite spin stiffness ${\cal D}_s (T>0)$ \cite{zotos96,hmeisner03,znidaric11}, the discovery of quasilocal charges \cite{prosen11,prosen13,prosen14,pereira14} have established lower bound for it. This finding fueled the efforts employing the thermodynamic Bethe Ansatz (TBA) approaches \cite{zotos99,urichuk19} and the framework of generalized Hydrodynamics (GHD) \cite{ilievski17,ilievski171,bulchandani18}. Both analytical approaches for XXZ model indicate on a discontinuous variation of ${\cal D}_s$ with the anisotropy $\Delta$ \cite{prosen11,prosen13}, most pronounced near the noninteracting limit $\Delta \to 0$. Numerical calculations reveal a smaller discontinuity 
 and this disagreement has not been yet resolved  \cite{steinigeweg14,bertini20}.
 Due to the larger Hilbert space, much less explored are the charge and spin stifnesses, ${\cal D}_{c,s}$, within the integrable 1D Hubbard model where in general Mazur inequality \cite{zotos97} and GHD \cite{ilievski171,fava2020} suggest finite ${\cal D}_c>0$ at $U >0$ away from half-filling \cite{karrasch17}.

In this work we present the method for calculation of stiffnesses in the integrable XXZ chain with Hamiltonian, $H=H_0+\Delta V$, and the Hubbard chain with $H=H_0+U V$ for small but nonzero interactions, $0<\Delta \ll 1$ and $0<U \ll 1$, respectively. Here, we assume that both $H_0$ and $H$ are integrable. We employ the observation that the reference systems of noninteracting fermions (NIF), as described by $H_0$, exhibit large degeneracies which exponentially increase with system size due to the specific dispersion relation of NIF, $\epsilon_k \propto \cos(k)$. The degeneracy is then lifted by the interaction $V$, which in the integrable case leads to a sudden decrease of the stiffnesses (to which we will refer as {\it the jump} further on). In particular it causes a decrease of the spin stiffness, ${\cal D}_s$, within the XXZ model, and both charge and spin stiffnesses, ${\cal D}_{c,s}$, in the 1D Hubbard model. In the Appendixes \ref{appa}  and \ref{appb}, we demonstrate that similar reasoning can be applied also to the case when $H_0$ describes interacting integrable systems with degenerate spectrum, e.g., the XXZ model at $\Delta=1/2$. 

The approach holds for any {\it extensive} observable $A$ in a system of size $L$. Namely, the stiffness can be defined as $D_A =T {\cal D}_A$ where $D_A = \langle \bar{A}\bar{A} \rangle/L$ and $\langle ... \rangle $ is the infinite-temperature averaging over the Hilbert space. Here, we introduce also averaging over infinite time window, \mbox{$\bar{A}=\lim_{\tau\rightarrow \infty}\int_0^{\tau}{\rm d}t\,{\rm e}^{iH_0t}A{\rm e}^{-iH_0t}/\tau$}. The latter averaging eliminates the matrix elements of the observable $ A$ which are outside of the degenerate subspaces of the Hamiltonian $H_0$,
\begin{equation}
\bar{A} = \sum_{m} A_{mm} |m \rangle \langle m| + \sum_{m \ne n: {\cal E}_m = {\cal E}_n } A_{mn} |m \rangle \langle n|, \label{sum1}
\end{equation}
where $A_{mn}=\langle m|A|n\rangle$ and $H_0|m\rangle ={\cal E}_m|m\rangle $. For clarity, we explicitly distinguish between the time-averaging $\bar{A}$, performed for the unperturbed Hamiltonian $H_0$ and time-averaging $\tilde{A}$ performed in the same way but for $H$. Consequently, the stiffnesses for the systems described by $H_0$ and $H$ equal $ \langle \bar{A}\bar{A} \rangle /L$ and $\langle \tilde{A}\tilde{A} \rangle/L $, respectively.

It is clear that the last term in Eq.~(\ref{sum1}) may contribute to $D_A$ only for models with degenerate spectra, as it is the case for NIF, and vanishes when the degeneracy is lifted. In order to account for lifting of the degeneracy, one can diagonalize the perturbation $V$ within {\it the degenerate spectrum} of $H_0$. Therefore, all relevant matrix elements of $V$ are contained in the time-averaged operator, $\bar{V}$. We solve the eigenproblem, $\bar{V} |\alpha\rangle =v_{\alpha} |\alpha\rangle$, and express the time-averaged quantities, $\bar{A}$, in the $\{|\alpha\rangle\}$ basis. Then, one can split $\bar{A}=\bar{A}^{\parallel}+\bar{A}^{\perp}$ into two orthogonal operators
\begin{eqnarray}
\bar{A}^{\parallel} &=& \sum_{\alpha,\beta:v_{\alpha}=v_{\beta}} (\bar{A})_{\alpha,\beta} |\alpha \rangle \langle \beta |, \\
\bar{A}^{\perp} &=& \sum_{\alpha,\beta:v_{\alpha} \ne v_{\beta}} (\bar{A})_{\alpha,\beta} |\alpha \rangle \langle \beta |.
\end{eqnarray}
When the degeneracy of $H_0$ is (at least partially) lifted by $\bar V$, only the former operator has nonzero matrix elements for states with equal energies. Consequently only $\bar{A}^{\parallel}$ contributes to the stiffness in the perturbed model. For all models considered in this work we have found that, up to high accuracy, the stiffness for weakly perturbed Hamiltonian equals
\begin{equation}
 \lim_{\Delta,U \to 0} \langle \tilde{A}\tilde{A} \rangle /L=\langle \bar{A}^{\parallel} \bar{A}^{\parallel} \rangle /L \label{main},
\end{equation} 
whereas the jump in the stiffness equals to $\langle \bar{A}^{\perp} \bar{A}^{\perp} \rangle /L$. The relevant question is to what extent the perturbative approach applies also for nonzero $\Delta$, when the energy levels start to cross. We demonstrate that due to integrability, it remains valid also beyond the level crossing. Therefore the integrability is essential for our approach, but the analysis is applicable to any other  quantum-lattice model near points with macroscopic degeneracies.
We stress that  the {\it macroscopic} degeneracies discussed in the present work should not be confused with the  pairs of eigenstates which in a real-valued Hamiltonian (i.e. without an external flux) are {\it doubly} degenerate 
and related to each other by  time reversal \cite{narozhny}.

Calculating the left-hand-side of Eq.~(\ref{main}) is a complex problem which involves diagonalization of the many-body Hamiltonian $H$. However, $\bar{V}$ is a block-diagonal matrix, thus the complexity of calculating the right-hand-side of Eq.~(\ref{main}) is determined by the dimension of the largest degenerate subspace of $H_0$.  Note that in the case when $H_0$ represents NIF, the degenerate subspaces as well as the explicit form of $\bar{V}$ can be (in principle) obtained via analytic calculations. Thus, Eq.~(\ref{main}) is an essential advantage for the numerical calculations. Namely, the diagonalization of degenerate subspaces is feasible up to {a} large system sizes, otherwise reachable only by approximate methods. It also allows to identify observables which commute (do not commute) with $\bar{V}$ and, consequently do not show (do show) jump when the degeneracy is lifted. 

\section{XXZ model}
We first consider the spin XXZ chain, \mbox{$H=H_0+\Delta V$}, with
\begin{equation}
H_0 = \frac{J}{2} \sum_{i} \left( c^\dagger_{i+1} c_i + \mathrm{H.c.}\right), \quad \quad 
V=J \sum_i n_{i+1} n_{i}. \label{xxz}
\end{equation}
Here we use the fermionic representation, periodic boundary condition (PBC) are assumed, $J=1$, and we express $\bar{V}$ using the momentum representation
\begin{equation}
\bar{V} = \frac{1}{L} \sum_{kpq} \epsilon_{q}\, \delta(\epsilon_{k+q} - \epsilon_{k} + \epsilon_{p-q} - \epsilon_{p})\, c^{\dagger}_{k+q} c_k c^{\dagger}_{p-q} c_p,
\label{v}
\end{equation}
where $\epsilon_k = \cos k$. The above equation yields (apart from a trivial choice $q=0$) two options: (1) $q = p-k$ when $\Delta \bar{V}$ represents the Hartree-Fock term, and (2) $p=\pi - k = \bar k$, i.e., $\epsilon_{p} +\epsilon_{k} =0$, for all $q$. The latter one is specific for the dispersion $\epsilon_k = \cos k$ and leads to large and nontrivial pairing degeneracy for NIF. This finding is crucial for further analysis and eventually for the jump in the investigated spin stiffnesses at $\Delta >0$. One may resolve the conservation of energy imposed by the $\delta$-function and find the explicit form of $\bar{V}$, 
\begin{equation}
\bar{V}= Q^{\dagger}_1 Q_1+\frac{1}{L} \sum_{k,p:k+p \ne \pi} [1-\cos(k-p)]n_k n_p. \label{vbar}
\end{equation}
We stress  that the crucial term $Q^{\dagger}_1 Q_1$ can be expressed as a product of local charges
which were introduced in Refs.~\cite{Fagotti2014,Essler2016}. They do not
conserve the particle number and are not translationally invariant  
\begin{equation}
Q_l = \frac{1}{\sqrt{L}} \sum_p e^{ipl} c_{\pi-p} c_p= \frac{1}{\sqrt{L}} \sum_i (-1)^i c_i c_{i+l}.
\end{equation}

At this stage we distinguish between two classes of local  conserved charges valid for the NIF system,
\begin{eqnarray}
E_m&=& \sum_k \cos(mk) n_k=\bar{E}_m, \nonumber \\ 
F_m&=& \sum_k \sin(mk) n_k=\bar{ F}_m . 
\label{EF}
\end{eqnarray}
One may show that $[Q_1,\bar{E}_{2m+1}]=0$ and $[Q_1, \bar{F}_{2m}]=0$. Consequently, $\bar{E}_{2m+1}=\bar{E}^{\parallel}_{2m+1}$ as well as $\bar{F}_{2m}=\bar{F}^{\parallel}_{2m}$ , and the corresponding stiffnesses do not exhibit a jump at $\Delta \to 0$. We note that these quantities correspond to the known (strictly local) charges in the XXZ model in the limit of $\Delta \to 0$, e.g., $F_2$ represents the energy current. On the other hand, $E_{2m}$ and $F_{2m+1}$ do not commute with $Q_1$ and the corresponding stiffnesses are expected to have a jump. In particular, it holds for the spin current $J_s =\bar{J}_s= F_1$.
Interestingly, the latter class corresponds to the symmetry sectors where stiffnesses are determined by the quasilocal charges \cite{ilievski15,mierzejewski15}.

If the jump $\langle J^{\perp}_s J^{\perp}_s \rangle /L$ remains nonzero for $L \to \infty $, then the operator $J^{\perp}_s$ is (at least) quasilocal, in view of the definition in Ref \cite{Ilievski16}. It order to show the latter property, we note that $\langle J^{\parallel}_s J^{\perp}_s \rangle=0$, hence $\langle J^{\perp}_s J^{\perp}_s \rangle =\langle (J_s-J^{\parallel}_s) J^{\perp}_s \rangle= \langle J_s J^{\perp}_s \rangle$. Then, one finds 
\begin{equation}
\lim_{L \to \infty} \frac{\langle J_s J^{\perp}_s \rangle^2}{L \langle J^{\perp}_s J^{\perp}_s \rangle } = \lim_{L \to \infty} \frac{\langle J^{\perp}_s J^{\perp}_s \rangle}{L} > 0,
\end{equation}
thus $J^{\perp}_s$ has nonzero projection on local and normalized operator $J_s/\sqrt{L}$ and is  quasilocal. Here, we recall that $J_s$ is an extensive operator hence its norm is $|| J_s||^2= \langle J_s J_s \rangle \propto L$. 

Next, we construct all many-body states for $N_f$ fermions. The studied basis is diagonal in the NIF occupations, $n_k$. It is crucial to single-out the pairs of states $p$ and $\bar p =\pi-p$ which are created by $Q^{\dagger}_1$. We distinguish between pair-states when $d_p = n_p+n_{\bar p} \in \{0,2 \}$, and singly occupied states with $d_k=1$
\begin{equation}
| n \rangle = \prod_{k} (c^{\dagger}_k)^{n_k} |0\rangle = \prod_{d_p=0,2} (a^{\dagger}_p)^{n_p}
\prod_{d_k=1} (c^{\dagger}_{\bar k})^{n_{\bar k}} (c^{\dagger}_{k})^{n_k} |0\rangle, \label{n}
\end{equation}
where $a^{\dagger}_p= c^{\dagger}_{\bar p} c^{\dagger}_p$. We consider systems with even $L$ where both $p, \bar p$ exist. Obviously, $N_f = 2N_d +N_s $, where $N_d,N_0,N_s$ are number of occupied pair-states, empty pair-states and singly occupied states, respectively. At fixed $N_f$, we have to deal with $N_{deg}=(N_d+N_0)!/(N_d ! N_0 !)$ degenerate states.

For even $L$ we can consider only $ |p| < \pi/2$, so that $|\bar p| > \pi/2$ (note that $|p| = \pi/2$ cannot form a pair), and the relevant perturbation can be written in terms of sums running {\it only} over $p,p' $ such that $d_p,d_{p'} \in \{0,2\}$, i.e.,
\begin{eqnarray}
{\bar V} &=& \frac{4}{L} \sum_{p \neq p' } \epsilon_{p'} \epsilon_p a^{\dagger}_{p'} a_{p} + \sum_p u_p n_p + \sum_{p \neq p'} w_{pp'} n_p n_{p'}, \label{vtilde} \\
u_p &=& \frac{4}{L} [ \epsilon_p^2 - \sin p \sum_{k \notin d} n_k \sin k], \quad w_{pp'} = - \frac{4}{L} \sin p ~\sin p'. \nonumber
\end{eqnarray}
The final  step is  then exact diagonalization (ED) of Eq.~(\ref{vtilde}) for each subspace with a fixed number $N_d,N_0$ of the pair-states. This restricts our numerical studies to $L \leq 28$, where the degeneracy can reach $N_{deg} \sim 3000$, but the dimension of the whole Hilbert space is $N_{st} \sim 10^8$. Here, we concentrate on (canonical) results obtained at fixed filling $n_{\mathrm av}=N_f/L =1/2$ and on the systems with $L = 4K$. The details of numerical calculations are discussed in the Appendix \ref{appa}. 

\begin{figure}[tb]
\includegraphics[width=1.0\columnwidth]{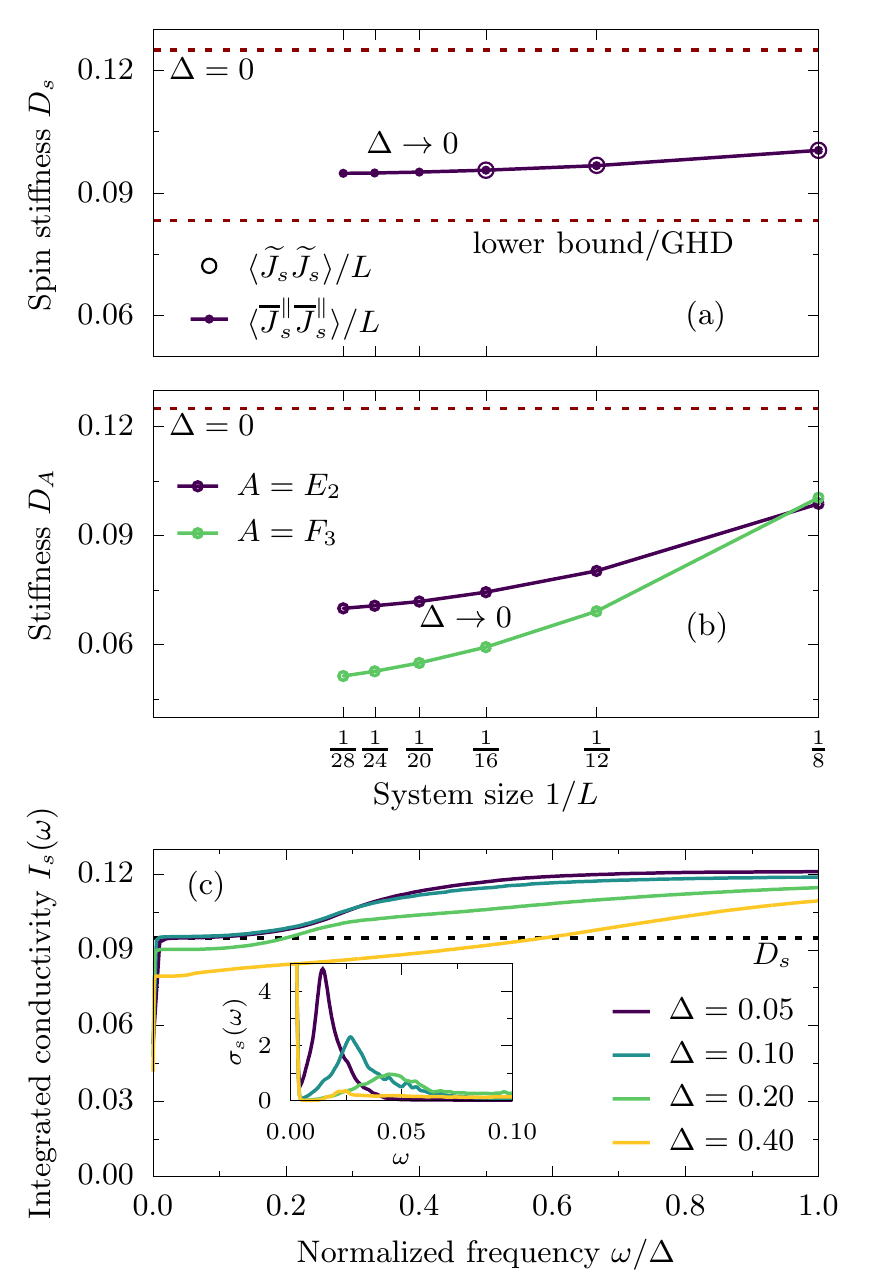}
\caption{XXZ model at half-filling $n_{\rm av}=1/2$. (a): Spin stiffness $D_s$ from the perturbative approach [right hand-side of Eq.~(\ref{main})] compared with (canonical) results from ED of the Hamiltonian [left hand-side of Eq.~(\ref{main})] for $\Delta= 0.01$. (b): Stiffness from the perturbative approach for the operators $E_2$ $F_3$, see Eq.~(\ref{EF}). In order to filter out the finite-size effects arising from $L$-dependence of the norm $\langle A A \rangle/L$, in (a) and (b) we show the renormalized stiffnesses, $D_A \times \langle A A \rangle_{L\to \infty}/\langle A A \rangle_{L}$. Results for $\Delta=0$  ($D^0_A = 1/8$) and analytical Mazur lower-bound/GHD result from Refs. \cite{prosen13,ilievski17}  are also shown. (c): Dynamical spin conductivity, $\sigma_s(\omega)$, and integrated conductivity, $I_s(\omega)$, obtained via MCLM on $L=28$ sites, respectively, in the inset and in the main panel, whereby $\omega$ is rescaled by $\Delta$ in the latter case. Horizontal line shows $D_s$ from (a).
} 
\label{fig1}
\end{figure}

The applicability of the perturbative approach for nonzero $\Delta$ is checked in two ways. First, we compare the stiffnesses which occur on both hand-sides of Eq.~(\ref{main}), whereby $\langle \tilde{A}\tilde{A} \rangle /L$ is ED result for the XXZ chain with $\Delta=0.01$. Fig.~\ref{fig1}a shows that the perturbative approach perfectly reproduces ED results, but it allows also to study much larger systems, when the $L$-dependence of $D_s$ becomes negligible. Calculated $D_s$ exceeds the Mazur lower-bound and GHD result in Refs.~\cite{prosen13,ilievski17} and is well consistent with numerical result obtained via typicality approach \cite{steinigeweg14,bertini20}. In Fig.~\ref{fig1}b we confirm our analytical predictions concerning the quantities for which corresponding stiffnesses $D_A$ show a jump at $\Delta \to 0$. In particular, $D_A$ for $E_2$ and $F_3$, see Eq.~(\ref{EF}), show a clear jump which is even more evident than for $J_s=F_1$. 

In order to identify  artifacts originating from improper order of limits, $L \to \infty$ and  $t \to \infty$, 
we have calculated the dynamical spin conductivity, $\sigma_s(\omega)$. We employ the microcanonical Lanczos method (MCLM) \cite{mclm} for $ L = 28$, where we reach the energy resolution $\delta \omega < 10^{-3}$. We present also the integrated conductivity $I_s(\omega) = \frac{2}{\pi}\int_0^{\omega} d\omega' \sigma_s(\omega')$ which allows to extract both the value of $D_s$ as well as the emergent incoherent part. The incoherent part is particularly important, as it contain the spectral weight which is eliminated from the stiffness, i.e., from the $\delta(\omega)$-peak, by lifting the degeneracy. When $\Delta$ increases, the incoherent peak shifts towards larger frequencies, as shown in the inset in Fig.~\ref{fig1}c. In the main panel we show the renormalized conductivity $I(\omega/\Delta)$. It confirms that the incoherent contribution scales linearly with $\Delta$, at least for weak $\Delta \leq 0.2$. Deviations become apparent only at substantial $\Delta \sim 0.4$. Also, $I_s(\omega \to 0) > 0.09$ at $\Delta \leq 0.05$. Both results  confirm the relevance of the perturbative approach for finite $\Delta > 0$. 
 
\section {Hubbard chain}
In analogy to the XXZ chain, we analyze the 1D Hubbard model, $H=H_0+U V$, with
\begin{equation}
H_0= -t_h \sum_{is}(c^{\dagger}_{i+1,s} c_{is} + H.c.) ,\quad V= \sum_i n_{i \uparrow} n_{i \downarrow}, \label{hub}
\end{equation}
where we assumed the PBC, set $t_h=1$, and consider $0<U \ll 1$. Again, one may find the time-averaged perturbation
\begin{eqnarray}
\bar{V}&=& Q^{\dagger}_0 Q_0 + \frac{1}{L}\sum_{k,p:k+p \ne \pi} 
(n_{k \uparrow}n_{p \downarrow} - c^{\dagger}_{k \uparrow}c_{k \downarrow} c^{\dagger}_{p \downarrow} c_ {p \uparrow} ) , \nonumber \\
Q_0 & = & \frac{1}{\sqrt{L}} \sum_p c_{\pi-p \downarrow} c_{p \uparrow} = \frac{1}{\sqrt{L}} 
\sum_j (-1)^j c_{j \downarrow} c_{j\uparrow}. \label{vhbar} 
\end{eqnarray}
The final step  consists in full ED of degeneracies. Here, we can study systems with $L \le 14$ sites with up to $N_{deg} < 4000$ and total $N_{st}\sim 10^8$ basis states, see the Appendix \ref{appa}.

It is clear that $\bar{V}$ in the Hubbard chain, Eq.~(\ref{vhbar}), closely resembles Eq.~(\ref{vbar}) for the XXZ model. However here, $Q_0$ annihilates spin singlets on a single lattice site. Again, $Q_0$ is a conserved charge of $H_0$ which does not conserve the particle number
and is not translationally invariant, while the $Q^{\dagger}_0 Q_0$ term is crucial for explaining the discontinuities of stiffnesses emerging for $U \to 0$. Performing additional summation over spin degrees of freedom in Eqs.~(\ref{EF}), one obtains {\it spin-symmetric}  conserved charges of $H_{0}$ analogous to $E_{2m+1}$ and $F_{2m}$ which {\it do commute} with $\bar{V}$ in the Hubbard chain. Such quantities coincide with local charges known for the Hubbard chain at $U \to 0$ \cite{grabowski95, zotos97}. In this limit known local charges commute with $ Q_0$ and their $D_A$ do not reveal a jump at $U \to 0$.

\begin{figure}[tb]
\includegraphics[width=1.0\columnwidth]{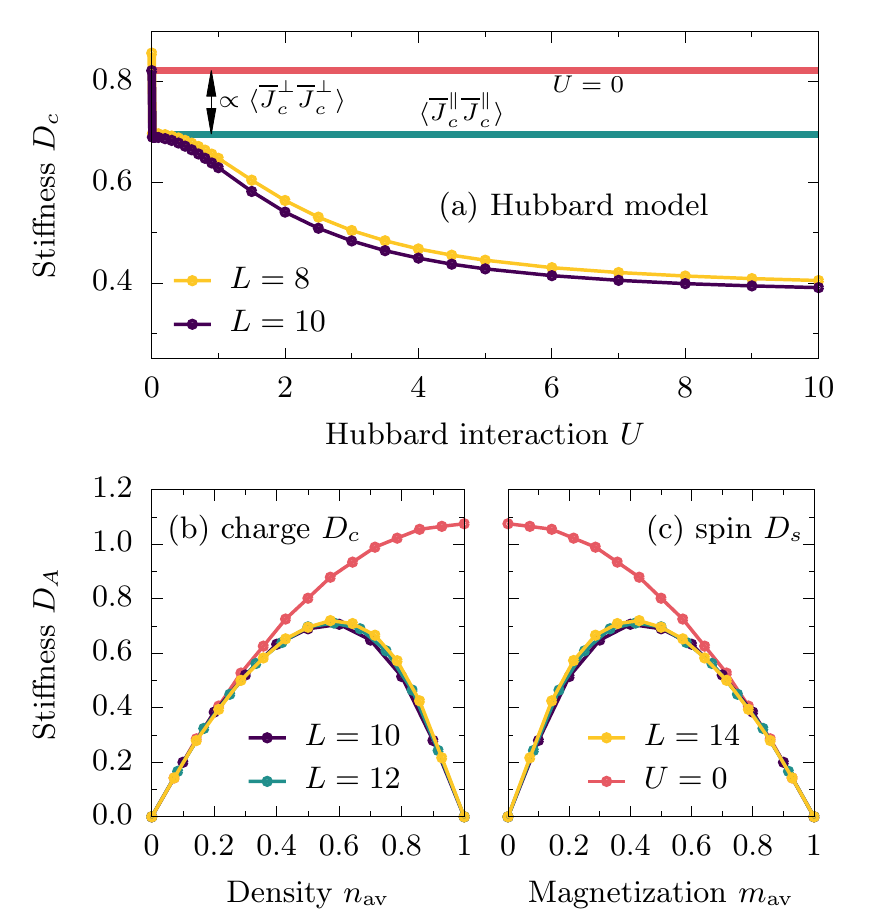}
\caption{Hubbard model. (a): Charge stiffness for $n_{\rm av}=1/2$ from ED, $ \langle \tilde{J_c}\tilde{J_c} \rangle /L$, (lines with point) and from perturbative approach, $ \langle \bar{J_c}^{\parallel} \bar{J_c}^{\parallel} \rangle /L$, (horizontal lines). (b): Charge stiffness, $D_c(n_{\rm av})$, for fixed  $m_{\rm av} \simeq 0$ and c): spin stiffness $D_s(m_{\rm av})$ for fixed  $n_{\rm av} \simeq 1$obtained from the perturbation approach. In (b) and (c) we show also $L=14$ results for $D_{c,s}^0$ at $U=0$.}
\label{fig2}
\end{figure}

A particular interest concerns the charge and spin currents
\begin{equation}
J_c = 2 \sum_{k,s} \sin(k) n_{ks}, \quad J_s = 2 \sum_{k,s} {\rm sgn(s)} \sin(k) n_{ks} ,
\end{equation} 
which do not commute with $Q_0$. Consequently the corresponding stiffnesses, $D_{c,s}$, are expected to exhibit a jump at $U \to 0$, depending on density $n_{\rm av}=(N_\uparrow+N_\downarrow)/L$ and magnetization $m_{\rm av}= (N_\uparrow-N_\downarrow)/L$. This is confirmed by numerical ED results for $D_c$ shown in Fig.~\ref{fig2}a at $n_{\rm av}=0.5$. We stress that the stiffness obtained from perturbation approach, $\langle \bar{J_{c}}^{\parallel} \bar{J_c}^{\parallel} \rangle /L=\langle J_{c}^{\parallel} J_c^{\parallel} \rangle /L$, very accurately reproduces the value obtained via ED of the many-body model, $ \langle \tilde{J_c}\tilde{J_c} \rangle /L$, see Eq.~(\ref{main}). 

Previous studies of the Hubbard chain focused on the half-filling, $n_{\rm av}=1$ and $m_{\rm av}=0$  \cite{karrasch17,ilievski17,ilievski171,bertini20}, where both stiffnesses were found to vanish. Our results confirm this finding and show that lifting degeneracy explains vanishing of $D_c$ (and more evident $D_s=0$). However, for intermediate $n_{\rm av}, m_{\rm av}$ results are also nontrivial and have been not presented so far. In Fig.~\ref{fig2}b we present two characteristic cases for $U \to 0$: $D_c(n_{\rm av})$ at fixed $m_{\rm av} \simeq 0$  (including also results for $m_{\rm av} =1/L$),  and $D_s(m_{\rm av})$ for $n_{\rm av} \simeq 1$ (allowing also for $ n_{\rm av}  =1+1/L $). One observes, that  $D_{c,s}$ remain in general large for  $U \to 0$ while the finite-size effects are rather insignificant. It should be also noted that $J_{c,s}$ have no overlap with known local charges, ${\cal I}_l$, in the Hubbard model for $U \to 0$, especially with the lowest local charge ${\cal I}_3$ \cite{zotos97}.  However, it holds true also for other ${\cal I}_{l>3}$ \cite{grabowski95}, for which $U \to 0$ are spin-symmetric generalizations of $E_{2m+1}$ and $F_{2m}$. 
On the other hand,  quasilocal charges plausibly responsible for $D_{c,s} >0$ have not been explicitly identified yet, although 
such charges should be accounted for within the GHD \cite{ilievski171,fava2020}. 
Due to similarity between Eqs.~(\ref{vbar}) and (\ref{vhbar}), we expect that $J_{c,s}^{\parallel}$ are quasilocal, as it is the case in the XXZ chain.

\section {Conclusions} 
It has previously been established that the spin stiffness, $D_s$, in the XXZ chain reveals a jump, when the many-body interaction, $\Delta$, is turned on. We have shown that this occurs as a manifestation of a general property of integrable systems with degenerate spectra and,  most importantly,  that the jumps may be explained accurately by perturbative lifting of degeneracy. 

We have demonstrated that these jumps occur in the XXZ chain at $\Delta \to 0$ as well as in the 1D Hubbard model for $U \to 0$ for a broad class of operators.
Such operators  do not commute with additional conserved charges which are specific for NIF with the nearest-neighbor hopping \cite{Fagotti2014,Essler2016}.
Interestingly, the jumps in XXZ chain occur in the symmetry sectors, where stiffnesses are known to originate from the presence of quasilocal conserved quantities. 

The same reasoning can be applied to other integrable lattice models near points of macroscopic degeneracies.
Results shown in the  Appendix \ref{appb} suggest analogous behavior for the XXZ chain with selected values $\Delta=\cos(\pi/n)$ with $n\in \mathbb{Z}$, although then the jumps are much smaller hence the quantitative interpretation of numerical results is less clear.
Perturbative lifting of the degeneracy allowed us also to calculate $D_s$ for large XXZ chains at $\Delta \to 0$, where we have found that $D_s$ is visibly larger than the known analytical  TBA/GHD  lower bounds. It is plausible that the non-commuting charges \cite{zadnik2016}  might be relevant for explaining the stiffnesses in the XXZ chains with $\Delta \to \cos(\pi/n)$, as  discussed  further on in the  Appendix \ref{appc}. Finally,  we have found that the known local charges  in the Hubbard chain are insufficient to explain the stiffnesses at $U \to 0$,  supporting the presence of quasilocal charges in this model.

\appendix
\section {Numerical calculations for the XXZ and Hubbard chains} 
\label{appa}
While the definition and treatment of degenerate NIF subspaces is described in the main text, the final step for each of $r$ different multiplets is the numerical exact diagonalization (ED) of Eq.~(\ref{vtilde}) which yields (in general) nondegenerate states $|\alpha \rangle$ with ${\cal E}_{\alpha} = E_r + \Delta {\cal E}_{rl}$. We turn now to the analysis of the spin current, expressed in NIF basis as $J_s = \sum_k \sin k (n_k - n_{\mathrm av})$. Subtracting $n_{\mathrm av}$ is useful at half-filling, $n_{\mathrm av}=1/2$, since then only $\pi$-pair states contribute, i.e., $J_s= \sum_p \sin p ~(2 n_p-1)$. It is in contrast to, e.g., the energy current $J_E = \sum_k \sin (2k) n_k$ where pair-state contribution vanishes and no discontinuity appears at $\Delta \to 0$. Of interest is the spin stiffness $D_s$, defined (at $T \to \infty$) as
\begin{equation}
D_s = \frac{1}{L N_{st}} \sum_{ \alpha,\beta: {\cal E}_\alpha = {\cal E}_{\beta}} |J_{\alpha\beta}|^2, \quad
J_{\alpha\beta} =\langle  \alpha  | J_s |\beta  \rangle,
\end{equation}
where $N_{st}$ is total number of many-body states, and we can consider here only the canonical average (at given $N_f$ and filling $n_{\mathrm av} =N_f/L$). We calculate $D_s$ as the sum of contributions over all degenerate multiplets  where all $J_{\alpha \beta}$ are evaluated and a small fraction of level may still remain degenerate ${\cal E}_\alpha={\cal E}_{\beta}$ even after ED. In this study we handle exactly the degenerate subspaces with up to $N_{deg} \sim 3000$ states, whereby the full summation over all NIF states $N_{st}$ has to be performed subsequently. This restricts our numerical results to $L \leq 28$, where $N_{st} \sim 10^8$. 

It should be recognized that, in general, the even-size systems $L=2 M$ with odd number of electrons $N_f$ can yield quite different (unphysical) result for $D_s$. In particular this appears at half-filling $N_f=L/2$ for $L \neq 4K$ where one gets anomalous result $D_s=0$ for $\Delta>0$, with the origin in additional degeneracies \cite{herbrych11}. We therefore present results for $D_s$ and its scaling with $1/L$ for canonical case and $L =4K$. One can partly avoid these problems by performing the full grand canonical (GC) calculation, but results still suffer from quite appreciable finite-size dependence in spite of $L \leq 28$. Much more stable results can be achieved by taking into GC sum only sectors with even $N_f$. 

\noindent {\it 1D Hubbard model.} 
Full diagonalization of degeneracies to the lowest order in $U$ follows the procedure analogous to the one described for XXZ model. Within the NIF many-body basis $| n \rangle = \prod_{k s} (c^{\dagger}_{k s} )^{n_{k s}} |0\rangle $, we determine first the degenerate multiplets having the same total momentum $q= \sum_{ks} k~ n_{ks}$. The multiplets emerge due to allowed spin flip and/or due to the $\pi$-pair occupied and empty states. The calculation is then performed within the canonical spaces, fixing $N_\uparrow, N_\downarrow$ for given $L$. By allowing up to $N_{deg} = 4000$ degenerate states within each multiplet we can reach $L \leq 14$ for arbitrary density $n_{\mathrm av}$ and magnetization $m_{\mathrm av}$, with up to $N_{st} < 10^8$ states.

\section{XXZ model for $\Delta=1/2$.}
\label{appb}

The energy spectra in the XXZ model are highly degenerate also for selected values of $\Delta=\Delta_c$. This degeneracy gives rise, e.g., to the presence of noncommuting quasilocal chargess derived in Ref. \cite{zadnik2016}. Moreover, the analytical lower bound on the spin stiffness in the XXZ chain reveals jumps in stiffnesses at $\Delta \to \Delta_c$ \cite{bertini20}. The latter jumps at $\Delta_c \ne 0$ are much smaller then those discussed in the main text for $\Delta \to 0$, making the numerical studies much more demanding. While we are not able to provide a reliable finite-size scaling towards $L \to \infty$, in this section we demonstrate for finite systems, that these jumps may be accurately described by the perturbative approach based on lifting the degeneracy. In particular, we discuss the case with the largest jump (apart from the case $\Delta=0$ ), i.e., we study the XXZ model for $\Delta \to \Delta_c=0.5$. 

The unperturbed Hamiltonian with degenerate spectrum reads
\begin{eqnarray}
H_0 &= &\sum_{i} \frac{1}{2}(c^{\dagger}_{i+1} c_i +{\rm H.c.}) + \Delta_c \sum_i n_{i+1} n_i, 
\end{eqnarray}
whereas the degeneracy is lifted in the full Hamiltonian for $\Delta \ne \Delta_c$
\begin{eqnarray}
H &= &H_0+ (\Delta-\Delta_c) V,\quad \quad V= \sum_i n_{i+1} n_i. 
\end{eqnarray}
In the following, we present numerical evidence that 
\begin{equation}
\lim_{\Delta \to \Delta_c} \langle \tilde{A}\tilde{A} \rangle /L=\langle \bar{A}^{\parallel} \bar{A}^{\parallel} \rangle /L \label{mainsup},
\end{equation} 
where the stiffnesses $ \langle \tilde{A}\tilde{A} \rangle /L$ and $\langle \bar{A}^{\parallel} \bar{A}^{\parallel} \rangle /L$ are defined in the same way as in the main text, but with appropriately modified Hamiltonians $H_0$ and $H$. For simplicity, here we apply GC averaging over the entire the Hilbert space, whereas in the main text we have used the canonical averaging over the subspace with a fixed number of particles.

\begin{figure}[!tb]
\includegraphics[width=0.8\columnwidth]{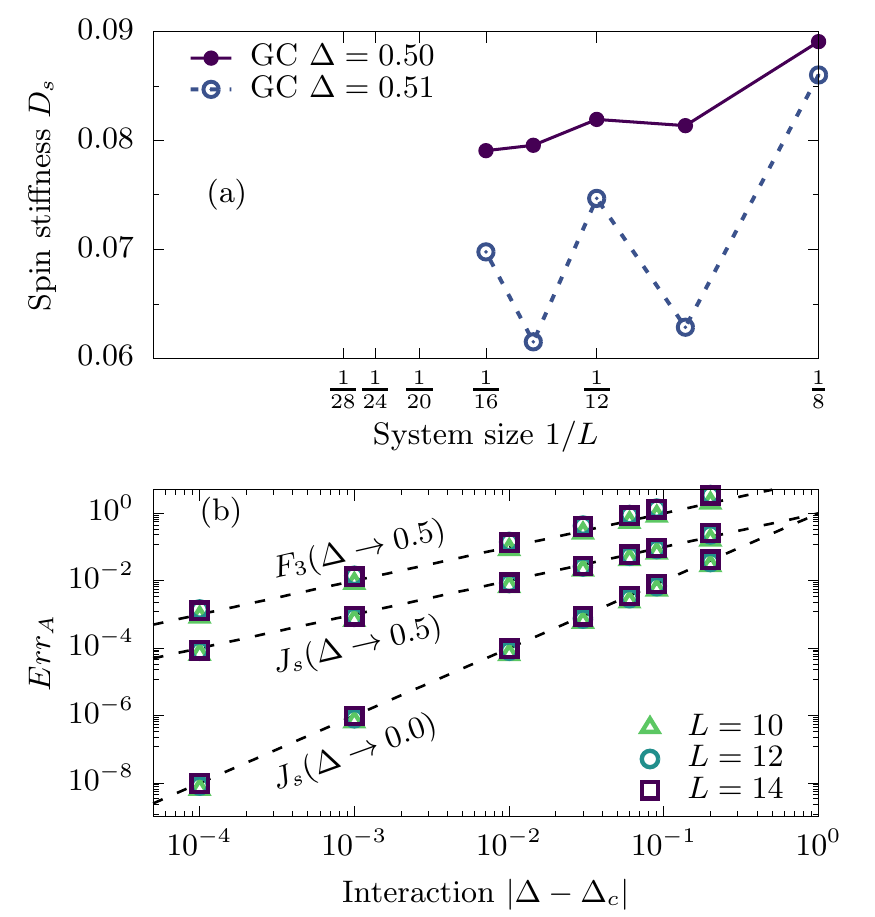}
\caption{Results for the XXZ chain with grandcanonical averaging over the Hilbert space. (a): Spin stiffness, $D_s$, for $\Delta=0.5$ and $\Delta=0.51$. (b) Relative difference,  $Err_{A}$, between the ED results and the perturbative approach, see Eq.~(\ref{erre}). We show results for $\Delta \to 0.5$ for the spin current $A=F_1=J_s$ and for $A=F_3$, see Eq.~(\ref{EF}) in the main text for the definition of operators. For clarity, the results for $F_3$ are multiplied by factor $10$. For comparison we show also the results for the spin current when $\Delta \to 0$.} 
\label{figS1}
\end{figure}

\begin{figure}[!htb]
\includegraphics[width=0.8\columnwidth]{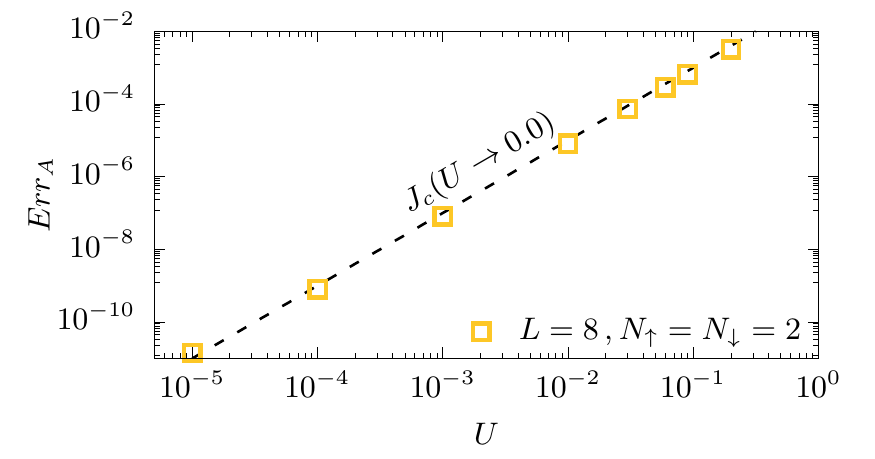}
\caption{ The same as in Fig.~\ref{figS1}b but for charge current ($A=J_c$) in the quarter-filled Hubbard chain. } 
\label{figS2}
\end{figure}

Figure~\ref{figS1} shows the spin stiffness, $D_s$, in the XXZ model obtained from ED of the Hamiltonian for $\Delta=\Delta_c$ and $\Delta=\Delta_c+0.01$. Since the difference between both results (i.e., the jump) is small while its $L$-dependence is quite significant, we are unable to judge whether the jump remains nonzero also in the thermodynamic limit. Nevertheless, based on the analytical lower bound on $D_s$ given in Ref.~\cite{bertini20}, we conjecture that it remains nonzero also for $L\to \infty$. In Fig.~\ref{figS1}b we show the relative difference between expressions on the both hand sides of Eq.~(\ref{mainsup}). Namely, we show 
\begin{equation}
Err_{A}=\frac{\langle \tilde{A}\tilde{A} \rangle- \langle \bar{A}^{\parallel} \bar{A}^{\parallel} \rangle }{\langle \tilde{A}\tilde{A} \rangle},
\label{erre}
\end{equation}
whereby $\langle \tilde{A}\tilde{A} \rangle$ is calculated with small but nonzero perturbation $|\Delta-\Delta_c|$ for $A=F_1=J_s$ and $A=F_3$, see Eq.~(\ref{EF}) in the main text. Results for the XXZ model with $\Delta \to \Delta_c$ are labeled as $\Delta \to 0.5$. For the sake of completeness, we show the same error obtained for the spin stiffness with the GC averaging for $\Delta \to 0$. Finally, in Fig.~\ref{figS2} we show the error (\ref{erre}) for the charge stiffness $D_c$  in the Hubbard chain for $U \to 0$. In the latter case, we have assumed the quarter filling  $n_{\rm av}=0.5, m_{\rm av}=0 $ and used the canonical averaging.

One may observe for all studied systems that the jump may be accurately described by the perturbative approach even for quite significant perturbations. It supports one of our main claims that lifting  degeneracy is a universal mechanism which explains the jumps of stiffnesses in various integrable models. 

\section{Products of noncommuting local charges and  the spin stiffness.}
\label{appc}

Numerical results shown in the main text in Fig.~\ref{fig1}a indicate that the  spin stiffness, $D_s$, in the XXZ model for $\Delta \to 0$ exceeds the analytical predictions obtained from the Mazur bound
 as well as the results from the generalized hydrodynamics (GHD).   The latter  is  unexpected, since the Drude weight obtained from the GHD  is widely believed to be exact in the thermodynamic limit. 
 Our results agree with other numerical studies  in Ref.  \cite{steinigeweg14,bertini20}.  We are not aware of any argument suggesting that  $L$-dependence of  $D_s$ changes above a certain length-scale  $L^*$, which is beyond the reach of  numerical approaches.   Moreover, Fig.~\ref{fig1}c does not indicate that there exists a low-frequency scale below $\omega < \omega^* \propto  \Delta$
which could lead to an irregular frequency-dependence of  the optical conductivity $\sigma(\omega \to 0)$,
so that our numerical results might be affected by incorrectly applied limits, $L\to \infty$ and $t \to \infty$.

The stiffness of an {\it extensive} operator $A$  is bounded by the Mazur inequality,
\begin{eqnarray}
D_A & \ge & \sum_{C}  \frac{1}{L} \frac{\langle A C \rangle^2}{\langle C C \rangle}.
\label{maz1}
\end{eqnarray}
where the summation runs over all orthogonal conserved charges, $C$. Typically, one considers only local and quasilocal charges for which \cite{Ilievski16} 
\begin{eqnarray}
\frac{1}{L} \frac{\langle A C \rangle^2}{\langle C C \rangle} \sim {\cal O}(1),
\end{eqnarray}
in the thermodynamic limit. In the main text, we have  introduced  additional local charges, $Q_n$,  which do not conserve the total $S^z$, 
are not translationally invariant and, in general, do not commute with charges which conserve $S^z$. 
It is known that analogous quasilocal charges exist non only for $\Delta=0$ but also for other commensurate values of 
$\Delta$  \cite{zadnik2016}. In other words, they exist exactly for these model parameters for which one obtains analytical results from the Mazur bound and the GHD. 
It is clear that for an observable $A$ which conserves total $S^z$, one obtains $\langle A  Q_n \rangle=0$. For this reason, such $Q_n$ have not been included in the analytical Mazur bound neither they are  accounted for in the GHD.

However, it is possible that for properly selected $Q_n$ and $Q_m$, one may introduce a Hermitian {\it nonlocal} conserved operator 
 \begin{eqnarray}
 X_{nm}=i Q^{\dagger}_n Q_m +{\rm H.c.},
 \label{xmn}
 \end{eqnarray}
such that
\begin{eqnarray}
\frac{1}{L} \frac{\langle A   X_{nm} \rangle^2}{\langle  X_{nm} X_{nm} \rangle} \sim {\cal O}(\frac{1}{L}).
\label{maz2} 
\end{eqnarray}
The $1/L$ dependence on the right hand side of Eq. (\ref{maz2}) comes from nonlocality of $ X_{nm}$. Such operators may contribute to the stiffness, 
$$\sum_{m,n} \frac{1}{L}  \frac{\langle A   X_{nm} \rangle^2}{\langle  X_{nm} X_{nm} \rangle}  \sim {\cal O}(1),$$
 provided that for local  $A$ there exists an extensive number of orthogonal $X_{mn}$ which fulfill Eq. (\ref{maz2}).  The general proof for commensurate $\Delta \ne 0$ goes far beyond the scope of the present work. 
However, for $\Delta =0$ one may rather easily show that an extensive number of  $X_{mn}$ satisfies   Eq. (\ref{maz2}) and the details are presented below.

In the main text we have studied the {\it noninteracting} limit of the XXZ model  and introduced local charges which do not conserve the total $S^z$  
\begin{eqnarray}
Q_n & = & \frac{1}{\sqrt{L}} \sum_j (-1)^j c_j c_{j+n} .
\label{qsy}
\end{eqnarray}
Beacuse of the  $1/{\sqrt{L}} $ factor,  these operators have $L$-independent norm, i.e., $\langle Q_n  Q^{\dagger}_n  \rangle  \sim 1$. 
We have also introduced a sequence of orthogonal  current-like operators, 
\begin{eqnarray}
F_n  & = &  \frac{i}{2} \sum_j  (c^{\dagger}_{j+n}  c_j -{\rm H.c.}) , 
\end{eqnarray}
whereby the first one represents the spin current,  $F_1=J_s$. In order to facilitate comparison with other studies, $F_n$ have been chosen as {\it extensive} operators with,  $\langle F_n  F_m  \rangle = \delta_{mn}
L/8$.
We study also $X_{mn}$ defined via Eqs. (\ref{xmn}) and (\ref{qsy}) and consider here {\it only} odd difference $m-n$. 

We use also a particle--hole transformation defined by 
\begin{equation}
U=(c^{\dagger}_L-c_L)(c^{\dagger}_{L-1}+c_{L-1}) \; ... \; (c^{\dagger}_2+c_2)(c^{\dagger}_1-c_1),
\end{equation}  
for which $U c_i U^{\dagger}=(-1)^i c^{\dagger}_i$. Direct calculations show the transformations of the studied operators
\begin{eqnarray}
U\;F_n U^{\dagger} &= &(-1)^n F_n, \\ 
U\;X_{nm} U^{\dagger} &= &X_{nm} -(i[Q_n,Q^{\dagger}_m]+{\rm H.c.}).
\end{eqnarray}
Then, one obtains the projection
\begin{eqnarray}
\langle J_s  X_{nm}  \rangle & = &\langle  U J_s U^\dagger U  X_{nm}  U^{\dagger} \rangle,  \nonumber \\
&=& -\langle J_s  X_{nm}  \rangle+\langle J_s   (i[Q_n,Q^{\dagger}_m]+{\rm H.c.}) \rangle. \nonumber \\
\label{proj}
\end{eqnarray}
For an odd difference, $m-n$, one finds the commutation relation
\begin{eqnarray}
i[Q_n,Q^{\dagger}_m]  & = & \frac{2}{L} \left[ F_{n-m}+(-1)^nF_{n+m} \right],
\label{com}
\end{eqnarray}
which may be used  in Eq. (\ref{proj}) yeildig
\begin{eqnarray}
\langle J_s  X_{mn}  \rangle & =  & \frac{2}{L}  \langle J_s \left[ F_{n-m}+(-1)^nF_{n+m} \right]   \rangle  \nonumber \\
          &=& \frac{1}{4}  ([\delta_{n-m,1}+(-1)^n\delta_{n+m,1}].
\end{eqnarray} 
Consequently for a single local operator, $J_s$, there exist an extensive ($\sim L$) number of nonclocal charges, e.g.,  $X_{m,m+1}$ for $m=1,2,3,...$ and ech of them has a projection 
\begin{eqnarray}
\frac{1}{L} \frac{\langle J_S  X_{m,m+1}  \rangle^2}{\langle X_{m,m+1} X_{m,m+1} \rangle} \sim \frac{1}{L}, 
\label{ost}
\end{eqnarray}
that scales as $1/L$.  The above calculations should not be considered as a proof that the difference between the numerical results and the GHD calculations originate from the presence of charges which do not 
conserve $S^z$ [to this end one needs to derive Eq. (\ref{ost}) for  commensurate $\Delta \ne 0$ and ensure that charges in the Mazur bound are mutually orthogonal]. Nevertheless, these calculations demonstrate that such origin of the discussed discrepancy cannot be excluded.  
We note,  that noncommuting  charges are present also in the integrable Hubbard chain, but only in the noninteracting limit, so the discussed scenario
does not apply to  the latter model.

\begin{acknowledgments}
The authors thank T. Prosen and E. Ilievski for fruitful discussions. M.M. acknowledges the support by the National Science Centre, Poland via project 2020/37/B/ST3/00020. J.H. acknowledges the support by the Polish National Agency of Academic Exchange (NAWA) under contract PPN/PPO/2018/1/00035. P.P. acknowledges the support by the project N1-0088 of the Slovenian Research Agency. The numerical calculation were partly carried out at the facilities of the Wroclaw Centre for Networking and Supercomputing.
\end{acknowledgments}
\bibliography{manujump}

\end{document}